**The Narrative Construction of Generative AI Efficacy by the Media: A Case Study of the Role of ChatGPT in Higher Education**

Yinan Sun*, Ali Ünlü**, Aditya Johri*

* Department of Information Sciences and Technology, *George Mason University*

** School of Education and Human Development, *University of Virginia*

**Abstract**

The societal role of technology, including artificial intelligence (AI), is often shaped by sociocultural narratives. This study examines how U.S. news media construct narratives about the efficacy of generative AI (GenAI), using ChatGPT in higher education as a case study. Grounded in Agenda Setting Theory, we analyzed 198 articles published between November 2022 and October 2024, employing LDA topic modeling and sentiment analysis. Our findings identify six key topics in the media discourse, with sentiment analysis revealing generally positive portrayals of ChatGPT's integration into higher education through policy, curriculum, teaching practices, collaborative decision-making, skill development, and human-centered learning. In contrast, media narratives express more negative sentiment regarding their impact on entry-level jobs and college admissions. This research highlights how media coverage can influence public perceptions of GenAI in education and provides actionable insights for policymakers, educators, and AI developers navigating its adoption and representation in public discourse.

**Keywords: ChatGPT, Generative AI, AI in Education, Media Discourse, Agenda Setting Theory**

# 1 Introduction

News media play a crucial role in shaping public and policymakers' perceptions and understanding of innovative and emerging technologies like ChatGPT, including their risks, benefits, and appropriate applications (Brossard, 2013; Ittefaq et al., 2024; Chuan et al., 2019). Since its launch in November 2022 by OpenAI, Chat Generative Pre-trained Transformer (ChatGPT) has quickly made headlines in the U.S. and around the world (Dempere et al., 2023; Meißner, 2024; Wang et al., 2023), particularly for its capacity to generate human-like content (e.g., images, text) in response to complex and varied prompts (e.g., languages, instructions, questions) through deep learning models (Dempere et al., 2023; Lim et al., 2023; Wang et al., 2023). This widespread attention has extended to higher education, where ongoing debates among faculty members, students, policymakers, and technologists discuss the role of ChatGPT and its appropriateness in institutional and academic settings, highlighting both opportunities and challenges (Gill et al., 2024; Lian et al., 2024; Wen et al., 2025). While these conversations occur within academic matters, news media can significantly influence public understanding and policymaking around the adoption of generative AI in higher education (Brossard, 2013; Chuan et al., 2019; Ittefaq et al., 2024). In this research, news media refers to printed news media covering a variety of domains, including local, regional, national, global, economic, business, medicine, health, and more, which shape public perceptions (Even et al., 2024).

Recent studies indicate that traditional news media, such as printed newspapers, still have significant influence, even in contexts where social media and other digital platforms dominate, such as the United Kingdom and the U.S. (e.g., Bruinsma et al., 2024; Ceron, 2014; Harder et al., 2017). Furthermore, compared to social media, video platforms, messaging applications, and online-only outlets, printed news media- particularly reputable national newspapers along with regional and local news sources- are viewed as more trustworthy and reliable (Deacon et al., 2024; Park et al., 2022). Topics that receive extensive coverage in traditional printed news media are perceived over time as more important by the public, while issues discussed on social media do not have the same effect (Bruinsma et al., 2024). Given this heightened attention, there is a pressing need for a deeper understanding of how ChatGPT's role in higher education is represented in the news media. Importantly, our research distinguishes between *media discourse* and *media narratives*. Media discourse refers to the language and interactions produced through print media that are public, planned, and on the record, shaping how information is framed and influencing how the public understands and responds to social, political, and technological issues. (O'Keeffe, 2011). Media narratives are specific storylines embedded within this discourse, depicting a sequence of events with a clear beginning, middle, and end, focused on emotional appeal, and causal effect relationships that make complex information more relatable and persuasive (Molina-Perez et al., 2024).

By analyzing both the broader media discourse and the specific narratives within it, this study explores how U.S. news media have portrayed the role of ChatGPT in higher education and influenced public sentiment surrounding it. Specifically, we collected 198 news articles published between November 2022 and October 2024 and applied Latent Dirichlet Allocation (LDA) topic modeling to identify main themes, alongside sentiment analysis using the SetFit framework. Grounded in the agenda-setting theory, our study provides an empirical basis for understanding how media discourse may influence public and policy-level perceptions of generative AI in higher education.

**2 Literature Review**

**2.1 ChatGPT and Higher Education**

The adoption of ChatGPT in higher education has gained significant traction (Hasanein & Sobaih, 2023; Freeman & Aoki, 2023; Strzelecki, 2024). Yet, its reception within academia remains contentious (Sok & Heng, 2024). Previous research indicates that ChatGPT can play both beneficial and detrimental roles in higher education (Božić & Poola, 2023; Wen et al., 2025). On the one hand, ChatGPT is seen as a valuable resource for delivering personalized learning experiences, enhancing accessibility, improving language skills, supporting research design and development, summarizing articles, boosting academic writing, preparing exams, facilitating problem-solving and data analysis, as well as assisting with grading and creating instructional content (e.g., Hasanein & Sobaih, 2023; Rahman et al., 2023; Sok & Heng, 2024). On the other hand, ChatGPT has raised various concerns, such as excessive dependence on technology, plagiarism and cheating, the potential for sensitive information leaks, inadequate expertise and authority, reduced social interaction, and the risk of incorrect or biased responses (e.g., Fuchs, 2023; Sok & Heng, 2024). Prior studies have provided valuable insights into how and to what extent ChatGPT can be utilized in higher education (e.g., Dempere et al., 2023; Fuchs, 2023; Sok & Heng, 2024). Nonetheless, no definitive conclusions have been reached regarding the appropriate role, the boundaries of its use, and responsible practices for teaching and learning.

**2.2 ChatGPT and News Media**

The public discourse, fostered by printed news media, influences how ChatGPT should be understood and applied across various domains, including higher education. For instance, Karanouh (2023) noted that Big Tech and its surrounding actors (e.g., OpenAI, Google, Microsoft) dominated the word frequency results observed in the corpus of mainstream media headlines from November 2022 to March 2023. Roe and Perkins (2023) reported that the primary themes regarding ChatGPT included 'impending danger,' 'explanation/ informative,' 'negative capabilities of AI or ChatGPT,' 'positive capabilities of AI or ChatGPT,' 'humorous/comedic,' and 'experimental reporting,' as reflected in UK news media headlines from January to May 2023. Meißner (2024) reported that the main themes related to ChatGPT were focused on automation or assistance in creativity, followed by support in educational contexts, enhanced quality of products and services, and increased efficiency, with the most frequently mentioned industries by education and teaching, as reported by major online news outlets in the German context between November 2022 and April 2023. Prior studies provided insights regarding how news media present the role of ChatGPT, suggesting its potential opportunities and challenges, including in higher education. Nonetheless, an in-depth understanding of how news media specifically conveys the role of ChatGPT in higher education is still limited (Mahmutović et al., 2024; Ryazanov et al., 2024; Tang & Chaw, 2024).

By the same token, previous research found mixed sentiment evidence regarding news media reporting on ChatGPT. On the one hand, Karanouh (2023) suggested that the findings from sentiment analysis regarding ChatGPT were perceived more positively than negatively in mainstream media. Meißner (2024) suggested that the news reports highlight opportunities more often than risks regarding ChatGPT in Germany. On the other hand, Ryazanov et al. (2024) observed that, compared to the overall positive sentiments from 1991 to 2018, public perception of generative AI has increasingly concentrated on potential threats and risks since the launch of ChatGPT. Roe and Perkin (2023) also noted that media representations of AI are often sensationalized and tend to emphasize warnings. Between these contrasting findings, Xian et al. (2024) identified that news media sentiment predominantly shows a neutral to positive stance, with business-related articles exhibiting a more positive tone, while regulation and security articles reflect a reserved, neutral to negative sentiment. These findings were valuable for understanding the broader sentiment towards news media about ChatGPT. However, the impact of this trend on the public's perception of ChatGPT's role in higher education remains unclear.

**2.3 News Media Coverage Regarding ChatGPT in Higher Education**

News media play a vital role in presenting the role of ChatGPT in higher education to the public (e.g., Akpan et al., 2025; Gesser-Edelsburg et al., 2024; Hanafi et al., 2025). For example, Tang and Chaw (2024) pointed out that discussions regarding ChatGPT in Malaysian education in newspaper articles highlighted its adoption as the most frequently discussed topic. Freeman and Aoki (2023) noted that dominant themes in media coverage of ChatGPT and education include the revolutionization of traditional teaching methods, ethical concerns, and the impact on grading systems. Additionally, Malaysian media generally emphasize ChatGPT's potential to transform education, whereas Japanese media focus on the bans against ChatGPT use in schools (Freeman & Aoki, 2023). Kikerpill and Siibak (2023) reported on how news media construct dominant social imaginaries of ChatGPT, highlighting its lack of accountability in the technology's release and its disruptive impact on the education sector. These findings are useful, as they provide an understanding of how news media shape public opinion regarding the role of ChatGPT in higher education in these countries. However, how the U.S. news media constructs narratives about ChatGPT remains underexplored, despite the U.S. being a key global reference point for higher education and AI innovation. To address this research gap, we aimed to respond to the following research questions:

- **RQ1**: What are the main topics regarding ChatGPT's role in higher education as presented in U.S. news media?
- **RQ2**: What are the main sentiments regarding ChatGPT's role in higher education as presented in U.S. news media?

## 3 Theoretical Framework

Agenda-setting theory provides a valuable perspective for examining the main topics and sentiments within the discourses of U.S. news media regarding ChatGPT's role in higher education. This theory posits that there is a relationship between the media and prioritized issues that deserve public attention (McCombs & Valenzuela, 2007). A key concept within agenda-setting theory is salience transfer, which refers to how the media establishes the importance and consideration of issues (Grzywinska, 2012). Through the selection, emphasis, and consistent reporting of specific topics, the media leads the public to regard certain issues as more significant (Wu & Coleman, 2009). Furthermore, agenda-setting theory consists of two levels (Chyi & McCombs, 2004). First-level agenda setting focuses on transferring object salience (Chyi & McCombs, 2004) through the frequency and prominence of media coverage, which directly impacts the degree of public attention and the relative importance that an issue receives (McCombs & Shaw, 1972; Coleman et al., 2009). This dimension is conceptualized to address RQ1, which examines the main topics in news coverage about ChatGPT in higher education. Second-level agenda setting expands the analysis to the characteristics assigned to these topics (Hester & Gibson, 2003). This level explores how an issue is discussed, including the tone, framing, and evaluative aspects of media coverage (Meraz, 2011; Ceron et al., 2016). RQ2 aligns with second-level agenda setting theory, as it investigates the sentiment connected to news coverage of ChatGPT's role in higher education. Considering that both the frequency and tone of news coverage can shape public discourse, using agenda-setting theory to understand how news media constructs the role of ChatGPT in higher education is appropriate.

## 4 Data and Methods

**4.1 Data Collection and Processing**

We collected news data from Nexis Uni, a platform that is the most referenced database in news content analyses (Buntain et al., 2023). Our search timeframe spanned from November 2022 to October 2024, coinciding with the initial release of ChatGPT by OpenAI. We focused our search terms on ("ChatGPT") AND ("higher education" OR "university" OR "college"). We limit our scope to the articles published in English that address the role of ChatGPT in U.S. higher education. The initial criteria led to a total of 3,313 news articles. To refine the dataset, we first removed exact duplicate articles with the same titles, resulting in 1,648 articles. We then removed near-duplicates that had identical content but different titles, publication dates, or publication venues, yielding 400 articles. Then, we excluded articles for two criteria: (1) there was no meaningful discussion regarding the role of ChatGPT in higher education, and (2) the role of ChatGPT in higher education was discussed, but it did not pertain to the U.S. higher education. The final dataset comprised 198 news articles from 95 news outlets, including their title, author, publication date, and publisher. All authors collaboratively developed and agreed upon the selection criteria for screening studies throughout the process.

**4.2 Data Profiling**

To provide an overview of the dataset, we categorized these news outlets into four groups: U.S. national news outlets, U.S. regional and local news outlets, university-affiliated news outlets, and business news outlets. These categories align with common classifications used in media research and are corroborated by existing literature (e.g., Shearer & Mitchell, 2021). Findings indicate that discussions about ChatGPT in higher education, as reported by the printed news media in our dataset, were predominantly driven by U.S. regional and local news outlets, with 89 articles (44.95%). This was followed by 56 articles (28.28%) from university-affiliated news outlets, 49 articles (24.75%) from U.S. national news outlets, and four articles (2.02%) from business news outlets.

## 4.3 Data Analysis

### 4.3.1 LDA Topic Modeling

Latent Dirichlet Allocation (LDA) is a generative probabilistic model that represents documents as random mixtures of latent topics, each defined by a distribution of words (Blei et al., 2003). The words with the highest probabilities in each topic typically indicate what the topic entails based on the word probabilities from LDA (Liu et al., 2019). To the best of our knowledge, very few studies have utilized LDA to evaluate how the role of ChatGPT in higher education was portrayed in U.S. news media, despite LDA's wide-ranging applications in exploring themes related to public discussions on ChatGPT (e.g., Lee, 2023; Wang et al., 2023; Ittefaq et al., 2025).

Before topic modeling, we first removed URLs, logos, digits, emojis, punctuation, stop words (e.g., "at, " "in," "her," "his," "last"), and irrelevant phrases (e.g., "Load-Date," "End of Document") from the news titles and text. Then, we tokenized and lemmatized each news item to its base or stem form.

To select the most suitable topic model, we adapted the approach of Gan and Qi (2021) and Gan et al. (2023), which suggests identifying the optimal number of topics by calculating coherence and perplexity scores based on the criteria of exclusivity, explainability, and repeatability. Coherence metrics represent the interpretability of the overall topic and are used to evaluate the quality of the topic (Gan & Qi, 2021). Perplexity refers to the degree of uncertainty when assigning documents to topics. One should aim for a model with reasonably high coherence and low perplexity scores to ensure both accurate predictions and interpretable topics (Blei et al., 2003). Using the open-source Python library Gensim (Islam, 2019), we compared models with 5 to 10 topics and selected the six-topic solution as optimal based on both metrics.

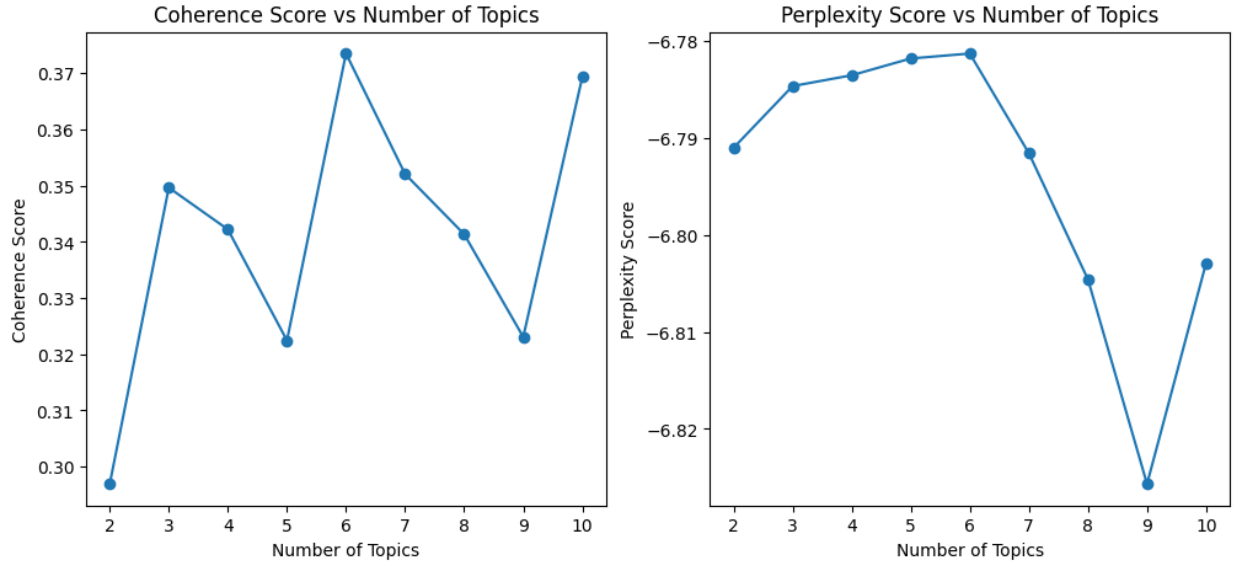

Fig. 1 Coherence and Perplexity Scores for Six Topics in LDA Modeling.

### 4.3.2 Sentimental Analysis

We conducted sentiment analysis on 198 news articles, totaling 88960 words and 10058 unique word forms. Given the size of our dataset, we selected the SetFit (Sentence Transformer Fine-Tuning) framework for sentiment analysis on our manually labeled dataset. SetFit is an efficient, prompt-free framework ideal for few-shot fine-tuning, particularly with small datasets, as it delivers high performance with minimal labeled data (Tunstall et al., 2022; Abdedaiem et al., 2023; Pannerselvam et al., 2024). Its ability to generalize effectively from limited annotations made it a perfect fit for our context, where large-scale supervised training was impractical (Pannerselvam et al., 2024).

We labeled 30 articles for sentiment analysis, designating 15 as negative (0) and 15 as positive (1), while the remaining 168 articles were left unlabeled. We utilized the pre-trained model "all-mpnet-base-v2, " which was fine-tuned with Cosine Similarity Loss over 100 iterations and five epochs, using a batch size of four. The model was trained on 70% of the labeled data and tested on the remaining 30%. The final model achieved an accuracy of 88.89% and an F1-score of 0.886.

After scoring each article, we conducted a monthly sentiment analysis of articles for each of the top six topics. The average sentiment scores were determined using a scale where positive sentiment was assigned a value of 1, and negative sentiment was assigned a value of -1. These values were then averaged monthly to produce scores ranging from -1 to 1 for normalized comparison.

**5 Results**

**5.1 Main Topics Regarding ChatGPT's Role in Higher Education in the U.S. News Media**

To explore the underlying themes, we extracted the top representative texts for each article within its assigned topic. Next, we conducted close reading and iterative thematic coding. Our analysis identified six distinct themes: 1) Jobs, Industry, and Young Workers; 2) AI/ML Skills and Career Aspirations; 3) Collaboration, Decision-making, and Bias; 4) College Admissions and Academic Integrity; 5) Policy, Curriculum, and Teaching Practices; and 6) Human-Centered Learning and Teaching.

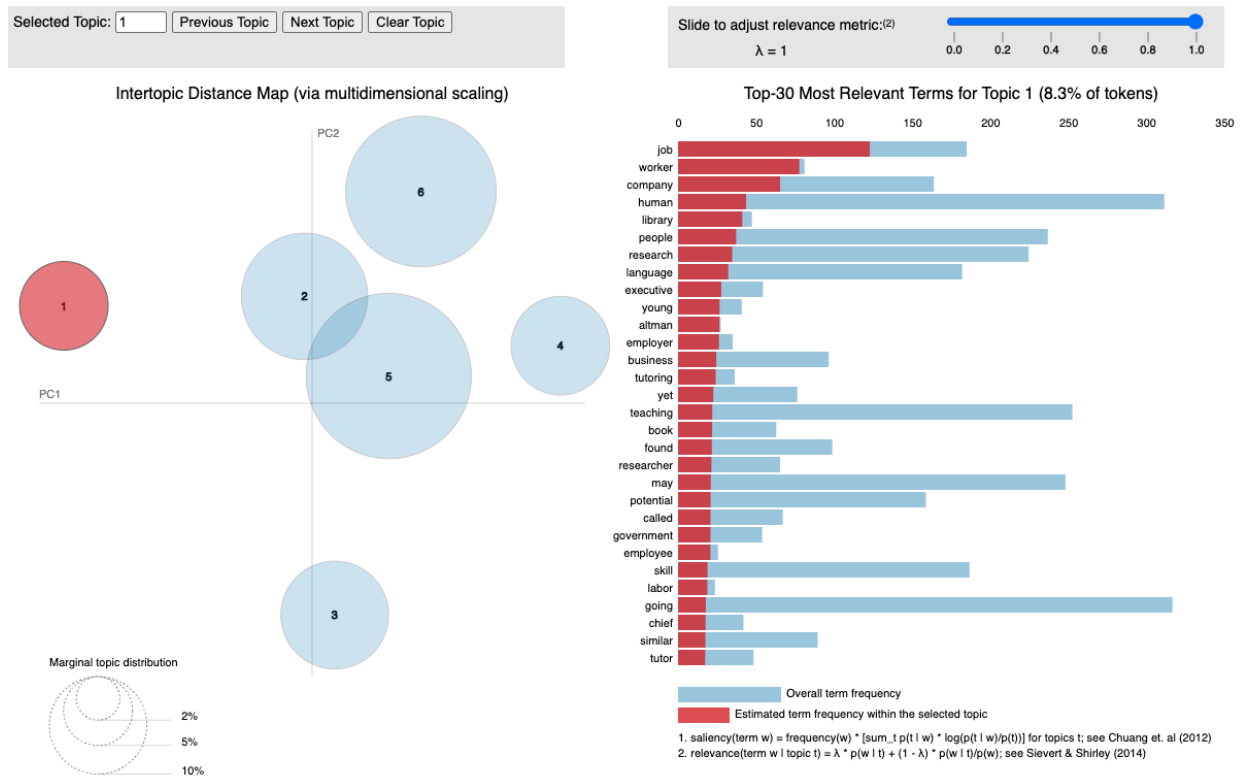

Fig. 2 Visualization of LDA model results with LDAvis.

Among the six identified topics, Topic 5 (Policy, Curriculum, and Teaching Practices) emerged as the most prominent, accounting for 77 articles (38.89%), followed by Topic 6 (Human-Centered Learning and Teaching) with 42 articles (21.21%). These two topics reflect prevalent media interest in how higher education institutions respond to the rapid integration of generative AI. Topic 2 (AI/ML Skills and Career Aspirations) was ranked third, featuring 29 articles (14.65%) that report how tools like ChatGPT have prompted universities to develop AI and machine learning-related degree programs while encouraging students to pursue careers in these fields. Topic 3 (Collaboration, Decision-Making, and Bias) was ranked fourth in prevalence, with 26 articles (13.13%), addressing the ethical implications of AI collaboration in areas such as institutional and community governance. Topic 4 (College Admissions and Academic Integrity) was cited in 14 articles (7.1%), while Topic 1 (Jobs, Industry, and Young Workers) was referenced in 10 articles (5.05%), making them the least frequently covered topics. Despite their lower frequency of coverage, both topics indicate a growing concern in media narratives: Topic 4 emphasizes fairness and cheating in admissions, while Topic 1 highlights job displacement and economic uncertainty among younger workers.

Notably, Topics 2 and 5 share overlapping themes because articles from both topics mention AI/ML-related courses and programs. However, the two topics diverge in focus. Topic 2 emphasizes how generative AI encourages universities to allocate resources to develop AI/ML-related programs and inspires students to pursue careers in these fields (e.g., McKinlay, 2023; Wethal, 2023), whereas Topic 5 focuses on how students, educators, and institutions react to the rapid integration of generative AI by offering training, workshops, and seminars (e.g., Paykamian, 2023; Singer, 2023c).

Specifically, Topic 1 explores how ChatGPT reshapes the job market, affecting companies, young professionals, and employers, touching upon issues such as automation, ethics, hiring practices, and young professionals. For example, Goldberg (2023) reported that while generative AI like ChatGPT may enhance productivity and create new job categories, it also risks exacerbating income inequality and job instability, particularly for white-collar, administrative roles, younger workers, and those in clerical fields. In contrast, Wiseman (2024) contended that while some fear AI would lead to mass layoffs, especially in roles such as customer service, writing, coding, and telemarketing, there was little evidence of widespread job loss. Focusing on young professionals, Carmichael (2023) noted that generative AI may impact entry-level job opportunities, especially for Gen Z individuals entering professional fields.

Topic 2 explores how the development of generative AI, such as ChatGPT, leads to the growing emphasis on AI and ML-related skills and career aspirations. For example, Wethal (2023) reported that UW-Madison built a 350,000-square-foot Computer, Data, and Information Sciences building. Similarly, Singer (2023a) reported that the University of Texas at Austin started a large-scale, low-cost online Master of Science degree program in artificial intelligence. McKinlay (2023) stated that Brigham Young University had added a machine learning degree to its undergraduate catalog.

Topic 3 focuses on the integration of generative AI in institutions and communities, collaborative decision-making processes, and the bias of AI systems, particularly in the field of health. For instance, Burke and O'brien (2023) reported that researchers at Stanford School of Medicine claimed chatbots were perpetuating racist, debunked medical ideas. Similarly, Collins (2023) noted that ChatGPT answered three dozen medication-related questions correctly or completely only a quarter of the time.

Topic 4 discusses the increasing impact of generative AI on college admissions and the associated concerns regarding academic integrity. For instance, Singer (2023b) reported that high school seniors were increasingly using generative AI tools like ChatGPT to prepare personal essays, though it remained unclear if college admissions offices were prepared for this trend. Schulten (2023) mentioned that the New York Times recently published two articles discussing how generative AI was disrupting essay writing, a process that has long been essential to the application process at elite colleges.

Topic 5 focuses on how educators and institutions respond to the opportunities and challenges presented by ChatGPT and other generative AI tools through revised policies, new teaching strategies, program development, and curriculum innovation. For instance, Singer (2023c) reported that some districts (e.g., Washington State) that once rushed to block generative AI are now attempting to embrace it. Paykamian (2023) noted that college professors across the U.S. found generative AI particularly useful for lesson planning and guiding classroom discussions.

Topic 6 examines the impact of generative AI on human-centered learning and teaching. For example, Magri-Nichol et al. (2023) noted that students at Colorado Mesa University believed that ChatGPT and AI language models may be useful tools, but they might also encourage laziness. Wilson (2023) reported that while ChatGPT can write everything from essays to poems and even computer code, good writing had no shortcuts as it requires revising, reflecting, and revising again.

**5.2 Main Sentiments Regarding ChatGPT's Role in Higher Education in the U.S. News Media.**

Our findings indicate that the overall sentiment regarding the role of ChatGPT in higher education is predominantly positive in the U.S. news media. Specifically, 65.2% of the articles (129 articles) displayed a positive sentiment, while 34.8% (69 articles) expressed a negative sentiment. The article with the lowest sentiment score (-0.8334) is " AI Could Perpetuate Racial Misconceptions in Health Care (Burke & O'Brien, 2023)," categorized under Topic 3 (Collaboration, Decision-making, and Bias). This article (Burke & O'Brien, 2023) highlights a Stanford-led study that found generative AI tools like ChatGPT may reproduce harmful racial biases and exacerbate existing disparities when responding to medical-related questions, adopting a critical and cautious tone. Conversely, the article with the highest sentiment score (0.8486) is "Discovering the Potential of ChatGPT and AI Tools at UNC Charlotte, " categorized under Topic 5 (Policy, Curriculum, and Teaching Practices). This article (*Inside UNC Charlotte,* 2023) reports that UNC Charlotte provides resources, sample syllabi, and workshops to promote the responsible use of generative AI tools like ChatGPT instead of banning them, showcasing a positive and optimistic tone.

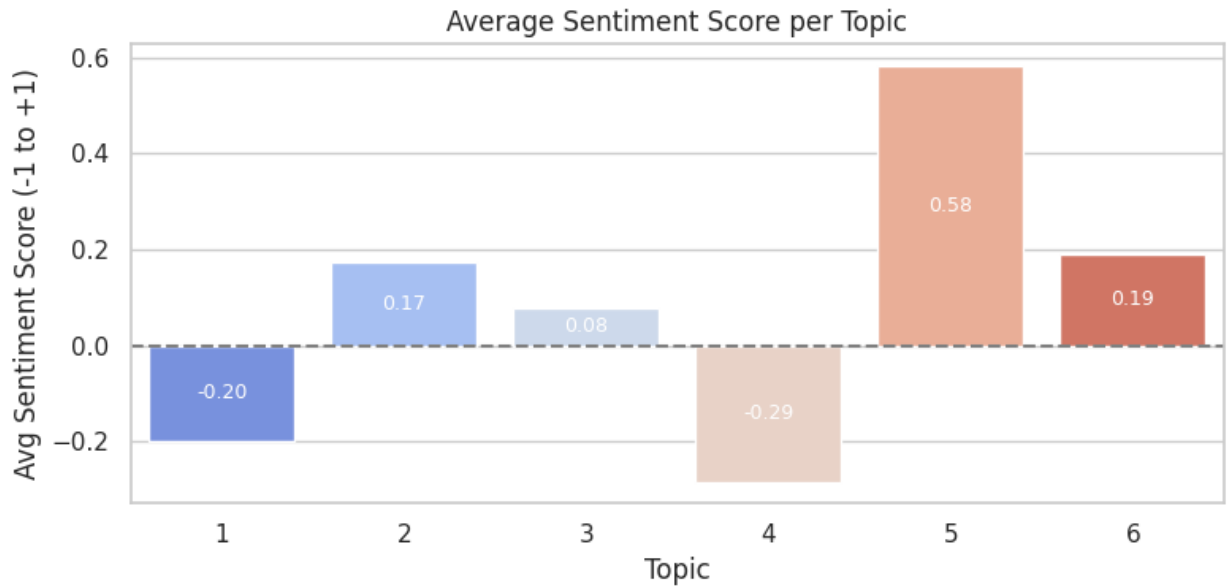

Fig. 3 Average Sentiment Score Across Topics.

The monthly sentiment analysis indicates that Topic 5 (Policy, Curriculum, and Teaching Practices) displays the most positive sentiment, reflecting favorable media attention toward institutional efforts to integrate generative AI into the higher education sector. Topics 2 (AI/ML Skills and Career Aspirations) and 6 (Human-Centered Learning and Teaching) demonstrated a moderately positive sentiment, suggesting cautious optimism about AI's potential to support skill development and enhance student learning experiences. Topic 3 (Collaboration, Decision-Making, and Bias) revealed a near-neutral but slightly positive sentiment, suggesting a blend of optimism and concerns surrounding the use of AI in institutional decision-making. In contrast, Topic 4 (College Admissions and Academic Integrity) exhibits the most negative sentiment, underscoring persistent ethical concerns about academic dishonesty, fairness, and the misuse of AI in the admission process. Topic 1 (Jobs, Industry, and Young Workers) also reflects a negative overall sentiment, highlighting concerns about labor displacement and economic disruption. These findings highlight the various aspects in which news media present generative AI in higher education in a nuanced way.

To be specific, Topic 5 (Policy, Curriculum, and Teaching Practices) demonstrated the most positive sentiment (0.58), highlighting universities' efforts to integrate generative AI into their practices, curriculum, and policy. Within Topic 5, the sentiment ranges from the lowest, "Multiple TAMU-C Students Exonerated after Being Suspected of Using AI to Cheat (Hairgrove, 2023)" (-0.8322), to the highest, "Discovering the Potential of ChatGPT and AI Tools at UNC Charlotte (Inside UNC Charlotte, 2023)" (0.8486) and "ChatGPT Sparks a Pivot in Academia: Why CT Professors Aren't that Concerned about Cheating via AI (Stannard, 2023)" (0.8486).

Topic 2 (AI/ML Skills and Career Aspirations) and Topic 6 (Human-Centered Learning and Teaching) exhibit a moderate positive sentiment, indicating optimism regarding AI/ML-related skills for career development and the influence of generative AI on human-centered learning. In Topic 2, the sentiment ranges from "How Student Defrauded Lehigh University International Teen Used Free Online Tools, Fake Emails to Obtain Full Ride (Pelekis, 2024)" (-0.8332) to "The University of Texas Plans Online A.I. Degree (Singer, 2023d) (0.8483). Similarly, the sentiment in Topic 6 spans from "How Well Can YU Faculty Detect ChatGPT (Levin, 2024)" (-0.8334) to "An Art Professor Says A.I. Is the Future. It's the Students Who Need Convincing (Small, 2023)" (0.8484).

Topic 3 (Collaboration, Decision-making, and Bias) has a slightly positive tone, indicating mixed feelings about integrating AI into the decision-making process while still addressing ethical concerns and biases. Within Topic 3, sentiments range from "AI Could Perpetuate Racial

Misconceptions in Health Care (Burk & O'Brien, 2023)" (-0.8334) to "At SFCC, Preparing Students for A Tech-Driven Future (Hood, 2024)" (0.8483).

In contrast, Topic 4 (College Admissions and Academic Integrity) shows the most negative sentiment (-0.29), indicating strong concern about the misuse of generative AI tools and fairness in higher education. Within Topic 4, the articles with the lowest sentiments include "We Used A.I. to Write Essays for Harvard, Yale and Princeton. Here's How It Went (Singer, 2023e)" (-0.8327), " Professors Ask: Are We Just Grading Robots (McMurtrie, 2024)" (-0.8323), and "'Guilty' of Cheating: An AI Cautionary Tale; Plagiarism Accusation Goes Genuinely Wrong (Jimenez, 2023)" (-0.8317) and the highest sentiment score "Faculty and Staff Adapt to Rise of AI-Generated Text Technology(Libow, 2023)" with (0.8478).

Topic 1 (Jobs, Industry, and Young Workers) also shows somewhat negative sentiment, revealing the anxiety over job displacement and young workers due to generative AI. Within Topic 1, the sentiment range spans from the lowest, "A.I.s Threat to Jobs Prompts Question of Who Protects Workers (Goldberg, 2023)" (-0.8332), to the highest, "Genai is Transforming Higher Education, and Florida is Leading the Way (Dobrin & Fraser, 2024)" (0.8476).

**6 Discussion**

This study examined the main topics and sentiments regarding the role of ChatGPT in higher education in U.S. news media. For RQ1, we identified six main topics: (1) Jobs, Industry, and Young Workers; (2) AI/ML Skills and Career Aspirations; (3) Collaboration, Decision-Making, and Bias; (4) College Admissions and Academic Integrity; (5) Policy, Curriculum, and Teaching Practices; and (6) Human-Centered Learning and Teaching. For RQ2, we found that among the six topics, Topic 5 showed the strongest positive sentiment, followed by Topics 6, 2, and 3. Topic 1 reflected a negative average sentiment, whereas Topic 4 exhibited the strongest negative sentiment

Our findings are partially in line with prior studies that reported main topics in news media regarding the role of ChatGPT in higher education, including academic integrity (Freeman & Aoki, 2023; Li et al., 2024; Nam & Bai, 2023; Tang & Chaw, 2023), limitations of AI capabilities (Li et al., 2024; Tang & Chaw) and workforce challenges (Li et al., 2024; Nam & Bai, 2023), the development of AI competence/literacy (Li et al., 2024; Tang & Chaw, 2023), practices for adopting AI for learning and teaching (Nam &Bai, 2023; Tang & Chaw, 2023), personalized learning (Freeman & Aoki, 2023), and the emphasis on critical thinking skills and social skills (Tang & Chaw, 2023). Furthermore, our analysis offered a more nuanced understanding with detailed insights. To elaborate, while generally aligned with prior studies that identified academic integrity as a key concern in generative AI discourse (e.g., Freeman & Aoki, 2023; Tang & Chaw, 2023), our analysis further extends prior research by highlighting that U.S. media narratives focus more specifically on the use of ChatGPT in college applications and the admissions process rather than on academic misconduct more broadly. The discourse regarding how AI might impact job markets is more centered on entry-level positions and challenges faced by young professionals rather than the overall job market. Additionally, U.S. news media tend to stress human agency when discussing how ChatGPT can facilitate human-centered learning and teaching instead of focusing on what the technology can do.

Agenda-setting theory posits that the more frequent and prominent the media coverage, the greater the attention and priority given to certain issues (McCombs & Shaw, 1972; Miller, 2005). Our findings indicate that when discussing ChatGPT's role in higher education, U.S. news media tend to emphasize how educators and institutions integrate it through policy updates, curriculum reform, and teaching practices, along with how ChatGPT can facilitate a human-centered education. It contrasts with media coverage in Malaysia and Japan, which more often emphasize the overall benefits and risks of generative AI (Tang & Chaw, 2023; Freeman & Aoki, 2023), and with mainstream STEM journals and magazines that stress institutional conflicts and crises resulting from the adoption of ChatGPT in academic settings (Nam & Bai, 2023). This suggests a more solution-oriented media narrative, emphasizing constructive integration and institutional adaptation rather than highlighting its risks or conflicts. From a first-level agenda-setting perspective, U.S. news media not only inform audiences about the existence of generative AI in higher education but also actively shape public perception of what matters most: policy reform, pedagogical opportunities, and the significance of human agency in adapting to technological change.

Our sentiment analysis reveals a predominantly positive media stance regarding the role of generative AI in higher education, which aligns more closely with recent research analyzing social media discourse about ChatGPT in education (e.g., Casillano, 2024; Li et al., 2024). Furthermore, our findings extend previous research by uncovering different aspects and degrees of positive sentiment, ranging from the strongest, such as adopting ChatGPT through policy and practices, facilitating a human-centered education, and developing AI and ML skills for career advancement, down to the minimum positive of integrating ChatGPT into the decision-making process. In contrast, our findings diverge from prior studies (e.g., Roe & Perkins, 2023; Ryazanov et al., 2024; Xian et al., 2024) that suggest news media, particularly in the Global North, tend to adopt a more critical or alarmist tone when reporting on ChatGPT and generative AI. This discrepancy can be partially explained by the domain-specific nature of higher education, where generative AI is often seen as an opportunity for innovation and reform rather than as a technological threat in the business or security sectors. From a second-level agenda-setting perspective, the positive tone observed in U.S. news media suggests an emerging sociotechnical imaginary of AI in education as a tool for enhancing learning, fostering skill development, and driving institutional transformation. Taken together, our findings demonstrate how U.S. news media have formed a constructive and positive narrative about ChatGPT's role in higher education, emphasizing adaptation and opportunity instead of disruption and risk.

## 6.1 Limitations and Future Work

This study recognizes several limitations that, in turn, highlight opportunities for further research. Firstly, our data collection was restricted to English-language articles. Considering the diverse media narratives in various linguistic and cultural contexts, future research could benefit from including non-English news articles to offer a more comprehensive and globally representative analysis of ChatGPT's portrayal in higher education.

Another limitation arises from our reliance on Nexis Uni's news database, covering the period from November 2022 to October 2024. To address this gap, future research could incorporate alternative databases, such as ProQuest TDM and Factiva, and extend the temporal scope.

The topic modeling approach used in this study presents two main challenges. First, the assumption that each document belongs to a single topic does not always align with the complex and multifaceted nature of news articles. Second, our analysis revealed instances where topics clustered news items from different thematic categories, likely due to LDA's reliance solely on textual similarity for topic identification.

To overcome these challenges, future research could leverage advanced topic modeling approaches such as BERTopic or other GPT-based models, which allow for multi-topic classification within a single news article. These techniques could provide a more granular and context-aware representation of topic distributions in media narratives.

Another key limitation pertains to the granularity of sentiment analysis. First, the SetFit model used in our study classifies sentiment into only positive and negative, excluding neutral sentiment. To address this limitation, we analyzed the average sentiment distribution of the six identified topics. Secondly, our current approach classifies sentiment at the article level, potentially overlooking variations in sentiment across different sections of a news piece. Future research could address this limitation by performing sentiment analysis at a more detailed level, such as paragraph-level or sentence-level classifications. Lastly, our analysis is based on a small dataset. This may limit the robustness and generalizability of our findings. Future studies could use a larger and more diverse dataset to improve the outcome.

**7 Conclusion**

This study investigates how U.S. news media have presented ChatGPT's role in higher education from November 2022 to October 2024. We identified six major topics, with predominantly positive media sentiment. Our findings contribute to understanding how the news media can influence public opinions and attitudes toward integrating ChatGPT and other generative AI tools in higher education. These insights offer valuable insights for policymakers, academic institutions, and technology developers aiming to navigate the integration of AI in educational settings more responsibly and effectively.

**Acknowledgments**

This work is partly supported by U.S. NSF Award# 2319137, 2112775, 1941186, 1939105, 1937950, USDA/NIFA Award# 2021-67021-35329. Any opinions, findings, and conclusions or

recommendations expressed in this material are those of the authors and do not necessarily reflect the views of the funding agencies.

Meraz, S. (2011). Using time series analysis to measure intermedia agenda-setting influence in traditional media and political blog networks. *Journalism & mass communication quarterly*, *88*(1), 176-194.

Nam, B. H., & Bai, Q. (2023). ChatGPT and its ethical implications for STEM research and higher education: A media discourse analysis. *International Journal of STEM Education*, *10*(1), 66.

O'keeffe, A. (2013). Media and discourse analysis. In *The Routledge handbook of discourse analysis* (pp. 441-454). Routledge.

Pannerselvam, K., Rajiakodi, S., Thavareesan, S., Thangasamy, S., & Ponnusamy, K. (2024, March). Setfit: A robust approach for offensive content detection in tamil-english code-mixed conversations using sentence transfer fine-tuning. In *Proceedings of the Fourth Workshop on Speech, Vision, and Language Technologies for Dravidian Languages* (pp. 35-42).

Park, S., Fisher, C., & Lee, J. Y. (2022). Regional news audiences' value perception of local news. *Journalism*, *23*(8), 1663–1681.

Paykamian, B. (2024, January 25). ASU partners with openai to enhance instruction and research. GovTech. https://www.govtech.com/education/higher-ed/asu-partners-with-openai-to-enhance-instruction-and-research

Pelekis, A. (2024, July 25). How Student Defrauded Lehigh University International Teen Used Free Online Tools, Fake Emails to Obtain Full Ride. The Morning Call. https://www.mcall.com/2024/07/25/how-an-international-student-was-able-to-defraud-lehigh/

Rahman, M. M., Terano, H. J., Rahman, M. N., Salamzadeh, A., & Rahaman, M. S. (2023). ChatGPT and academic research: A review and recommendations based on practical examples. *Rahman, M., Terano, HJR, Rahman, N., Salamzadeh, A., Rahaman, S.(2023).*
30